# An Improved Body Shape Definition for Acoustic Guitars

Mark French

Purdue University

121 Knoy Hall

401 N. Grant St

West Lafayette IN 47907

guitar@purdue.edu

**Abstract**

Acoustic guitar body shapes usually belong to one of a small number of families of body shapes.  However, these shapes are not standardized or even precisely described in the literature on the subject.  Rather, they are the result of accumulated tradition and shapes vary so much that many common components must be treated as effectively custom parts.  Conventional curve fits are not possible because the shape is not a single valued function.  Numerical descriptions such as spline fits will work, but the resulting data is too cumbersome to be easily portable and may be dependent on choice of software.  Transforming the problem from rectangular to polar coordinates allows the use of a closed form expression to describe a family of body shapes in a compact and unambiguous way that is easy to implement in widely available software.

**Background**

There are a small number of commonly-used acoustic guitar body shapes.  However, there doesn't appear to be clear, compact mathematical descriptions of them; even widely used references often describe shapes only in general terms [1,2].  If guitars are to be made using computer controlled tools such as CNC routers, it makes sense to establish compact, unambiguous definitions of body shapes.

Figure 1 shows a family of acoustic guitar designs from Bourgeois Guitars that are representative of the body shapes currently in use.  Naming conventions for acoustic body shapes are, like much else in the guitar world, not standardized.  Probably the most common terminology stems from models produced by Martin and is reflected in the Bourgeois designs.

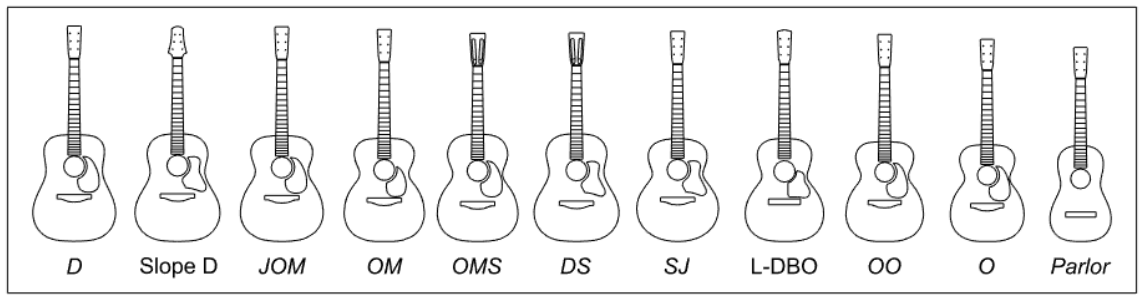

**Figure 1.  Some Common Guitar Body Shapes (bourgeoisguitars.net, image used with permission).**

It is worth noting that some manufacturers, including Taylor Guitars, use a naming convention that goes, in order of increasing size: mini, parlor, grand concert, grand auditorium, dreadnaught, jumbo.  There is significant overlap between the two families of designs, but the Martin convention seems to be most widely accepted.  That said, the method presented here is general and doesn't limit the user to one family or another.

In a world where computer-controlled machine tools are used to make guitars in environments ranging from large factories to home shops, it is especially important to have precise, mathematical definitions of body shapes.  Small builders often use full size plans that are, in turn, copied from existing instruments. Alternatively, some luthiers work from proportions developed from groups of existing guitars. Figure 2 shows proportions for classical guitars as suggested by Richard Bruné [3], an accomplished player and builder.

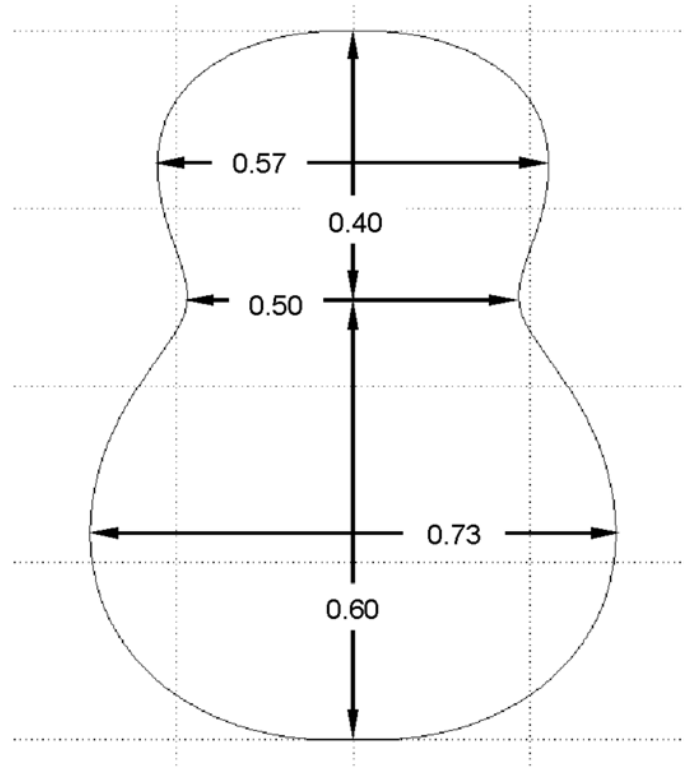

**Figure 2. Classical Guitar Proportions.**

There is essentially a hierarchy of precision in defining guitar body shapes. At the lowest level is matching a shape to a printed paper plan or drawing a plan based on a list of general dimensions. While these approaches certainly work for individual builders, they are empirical and not appropriate for production. Mathematical definitions of body shapes generally take one of three forms:

1) Drawing based on a few basic dimensions, often expressed as intersections of line segments, circles and ellipses.

2) Collections of curves (splines) based on drawings or basic dimensions

3) Closed form expressions

Using Computer Aided Design (CAD) software, it is easy to define complex curves using splines, a succession of short, simple curves that are stitched together to make the final shape. A common form is Non-uniform Rational B-Splines or NURBS. In practice, NURBS are often cubic polynomials [4].

Splines have the advantage that they are mathematically precise and unambiguous. However, it can be awkward to move spline data around. Rather than a few coefficients that might be used to define a function, splines require a data file containing lists of coefficients. There is also always the risk that different software packages might store and import spline data differently. Often, 2D shapes are passed as vector files in which a large collection of points are connected by short line segments. Even if well-implemented, it fundamentally misses the point of rigorously defining a body shape since it is simply storing a picture of the desired shape rather than working from instructions on how to make it.

An attractive alternative is to use a closed form mathematical expression – one that can be simply written out and evaluated using some kind of generic calculation software like Excel, MATLAB or OCTAVE. Once a suitable function has been defined, it is independent of whatever software is used to implement it and can be written out completely in a very compact form. Thus it is easy to share and hard to misinterpret.

**Polar Curve Fit**

The next step is to find an expression that fit those points precisely and smoothly. The work presented here is an improved implementation of an approach published earlier [5]. The previous approach resulted in small irregularities that limited their utility. Figure 3 shows the discrete points used for the curve fit. The example shown here began with a small number of points I measured from an existing guitar and modified slightly to suit aesthetic preferences. Since the body is symmetric about the center line, it is only necessary to model one side. Please note that the guitar industry in the United States still uses decimal inches almost exclusively, so the dimensions here conform. The method described here is independent of units.

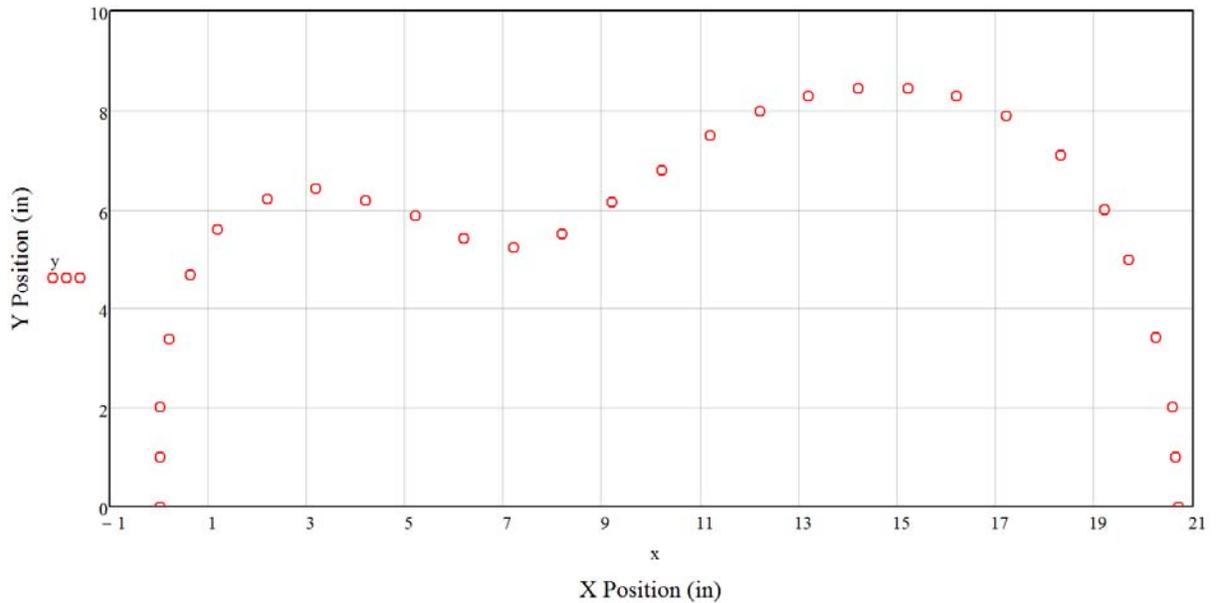

**Figure 3. Points Chosen for Guitar Body.**

Common forms of curve fits, including those implemented in popular suites of office software, assume single valued functions and cannot accommodate vertical slopes. However, both of these requirements are violated by the shape of acoustic guitar bodies. As shown in Figure 3, the slopes are vertical at both ends of the body. Rotating the body 90° makes the problem worse since there will be a vertical slope at three points and the shape is also not single valued. Fortunately, there is a simple way to fix both problems.

Transforming the problem into polar coordinates can give a curve that is both single valued everywhere and free of vertical slopes. Since the location of the origin of the coordinate system is arbitrary, it can be chosen to avoid numerical conditioning problems. For this body shape, placing the origin at x=9, y=0 gave a shape that posed no inherent problems, as shown in Figure 4.

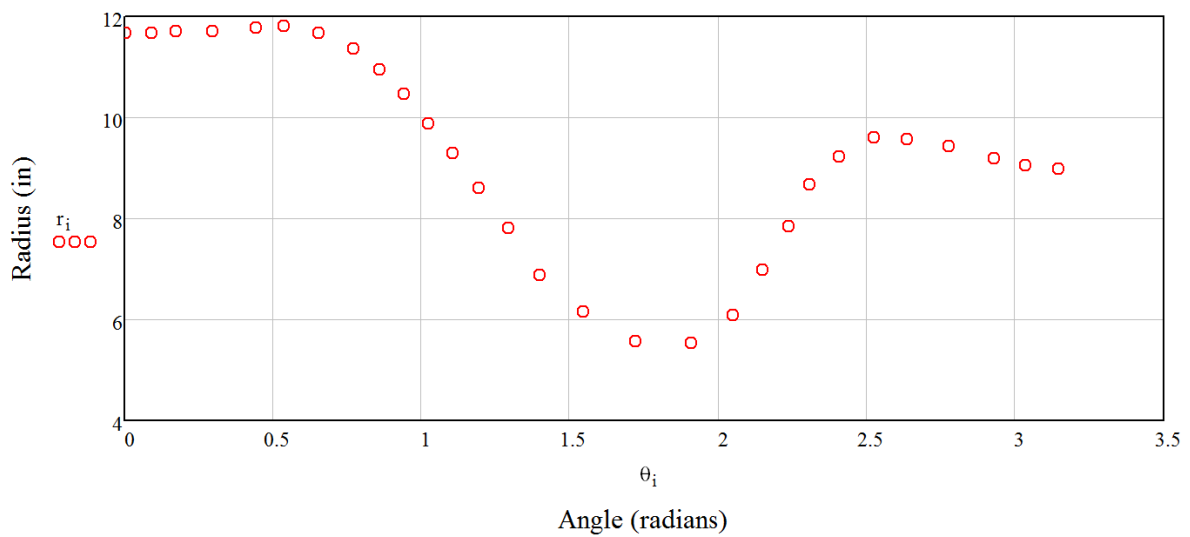

**Figure 4. Polar Coordinates from Guitar Body Plotted on Rectangular Axes.**

Fitting the points is now a much more tractable problem. The process progresses in two steps. The first is to get close to the points with a general purpose function and the second is to add some refinements to make sure resulting function is practical for making guitars.

The most useful class of general functions identified so far for this problem is rational polynomials. In the initial implementation, the shape was approximated by a 9/9 order rational polynomial without modification. While it worked well enough to demonstrate the method, it had some small scale deviations that were obvious when using a CNC router to make forms. A refinement uses a 5/5 order function with two small correction terms and the result is now being used for making a family of guitars. The parameters for the rational polynomial were identified using the MATLAB curve fitting tool [6]. The form of the function is

$$y(\theta) = \frac{p_1\theta^5 + p_2\theta^4 + p_3\theta^3 + p_4\theta^2 + p_5\theta + p_6}{\theta^5 + q_1\theta^4 + q_2\theta^3 + q_3\theta^2 + q_4\theta + p_5} \tag{1}$$

Figure 5 shows the 5/5 order curve fit superimposed over the points used to create it. At first glance, it may look acceptable, perhaps enough to be used for making guitars. However, a closer look shows that there are lingering differences that can cause practical problems when manufacturing guitar bodies.

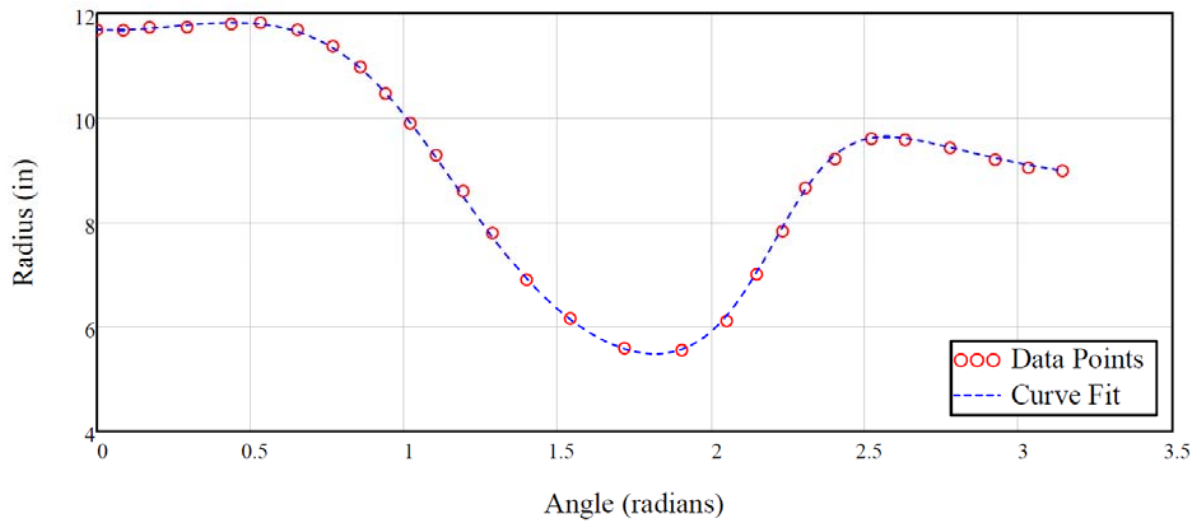

**Figure 5.  A 5/5 Order Curve Fit Superimposed over Points Used to Create It.**

A practical problem was that the function describing the body form is not straight where the neck would connect, as shown in Figure 6. This is where θ is close to π (3.14159 radians).

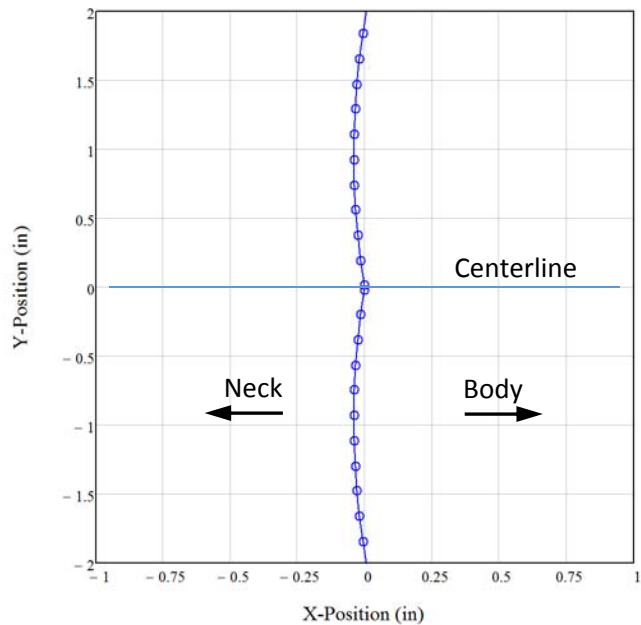

**Figure 6. Uncorrected Body Shape at Neck Joint.**

The simplest way to fix this problem is to add a correction term. This one needs to be close to zero everywhere except where θ is close to π (close to 180°). A good correction term is

$$\Delta_1(\theta) = a_1 e^{a_2(\theta - a_3)} \tag{2}$$

This is nearly zero everywhere except the region close to the neck joint as shown in Figure 7.

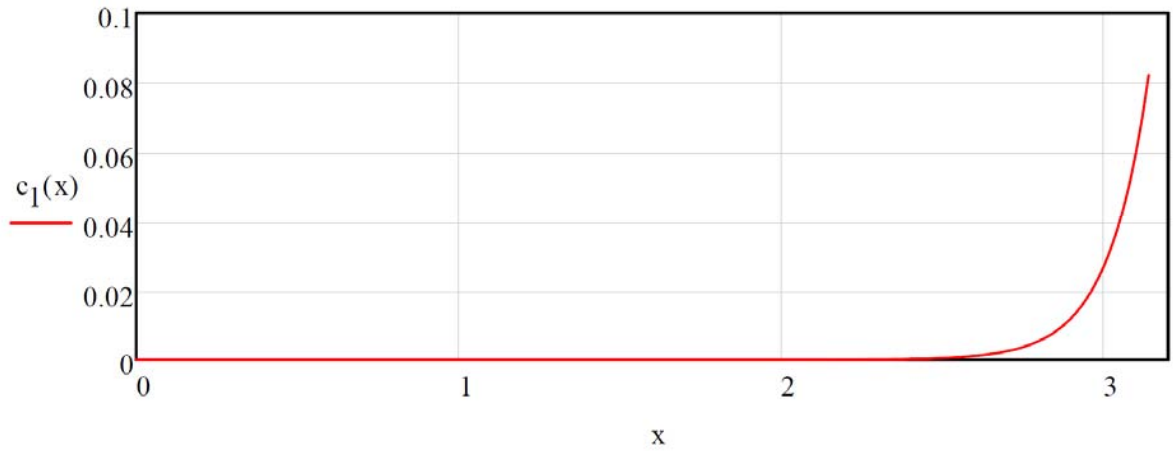

**Figure 7. Shape Correction Function.**

The body shape is now flat and vertical at the neck joint, as shown in Figure 8. Note that the correction function (red diamonds) is magnified by a factor of 2 in order to make it easier to see.

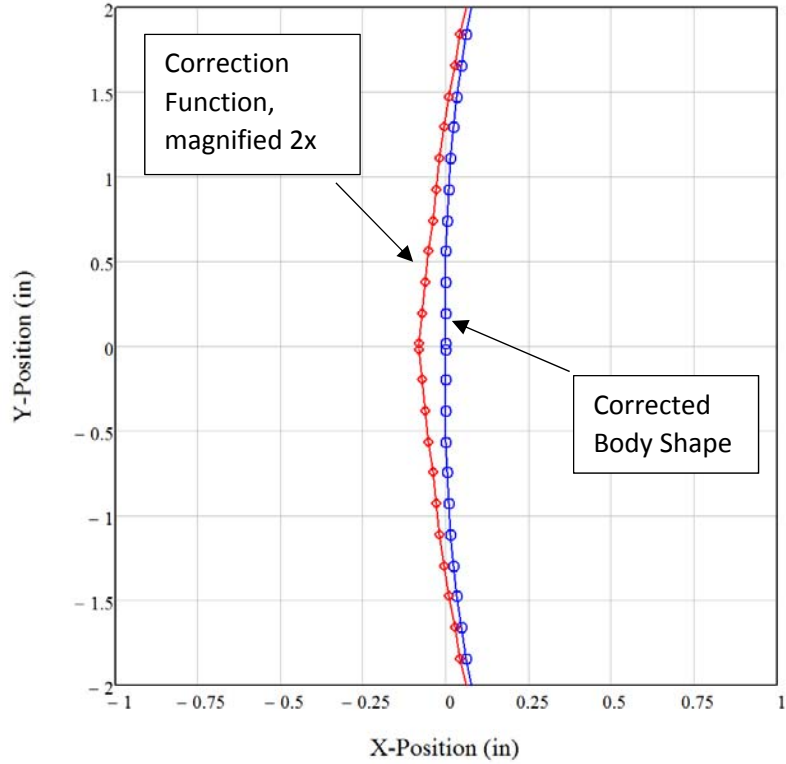

**Figure 8. Result of Neck Correction.**

A similar correction function makes sure that the slope at the tail end of the body is also vertical.

$$\Delta_2(\theta) = b_1 e^{-b_2 \theta} \tag{3}$$

**Resulting Body Shape**

Figure 9 shows the resulting body shape with both correction terms after transforming back into rectangular coordinates. It is largely free of the small scale variations that can cause problems with CNC machining.

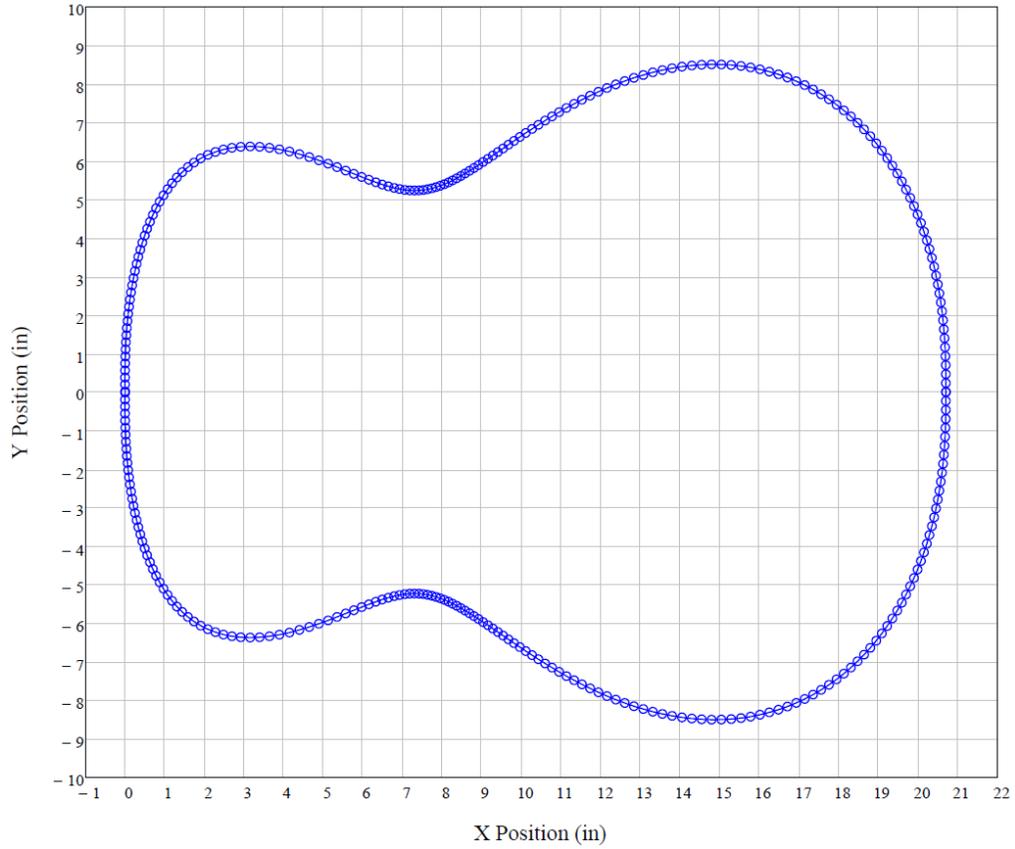

**Figure 9. Final Body Shape.**

The final body shape, with correction terms is

$$y(\theta) = \frac{p_1\theta^5 + p_2\theta^4 + p_3\theta^3 + p_4\theta^2 + p_5\theta + p_6}{\theta^5 + q_1\theta^4 + q_2\theta^3 + q_3\theta^2 + q_4\theta + p_5} + a_1 e^{a_2(\theta - a_3)} + b_1 e^{-b_2\theta} \qquad (4)$$

And the parameters are listed in Table 1.

**Table 1. Parameters for Body Curve Fit.**

| Parameter | Value |
|---|---|
| $p_1$ | 8.933 |
| $p_2$ | -60.03 |
| $p_3$ | 151.9 |
| $p_4$ | -167.31 |
| $p_5$ | 54.5 |
| $p_6$ | 23.51 |
| $q_1$ | -6.449 |
| $q_2$ | 15.26 |
| $q_3$ | -15.45 |
| $q_4$ | 4.772 |
| $q_5$ | 2.01 |
| $a_1$ | 103 |
| $a_2$ | 8 |
| $a_3$ | 4 |
| $b_1$ | -0.08 |
| $b_2$ | -8 |

The body length is 20.7 inches (about the maximum for practical instruments) and the lower bout width is 17 inches. However, it is easy to scale the x and y dimensions to give the size desired. Figure 10 shows the original body shape along with the shape scaled for a parlor guitar and a ¾ size travel guitar – approximately the body dimensions of a Baby Taylor. Another option is to divide the list of (x,y) coordinates for the body by 20.7 so that the body length is 1. Then, the body can be scaled by simply multiplying the coordinates by the desired body length.

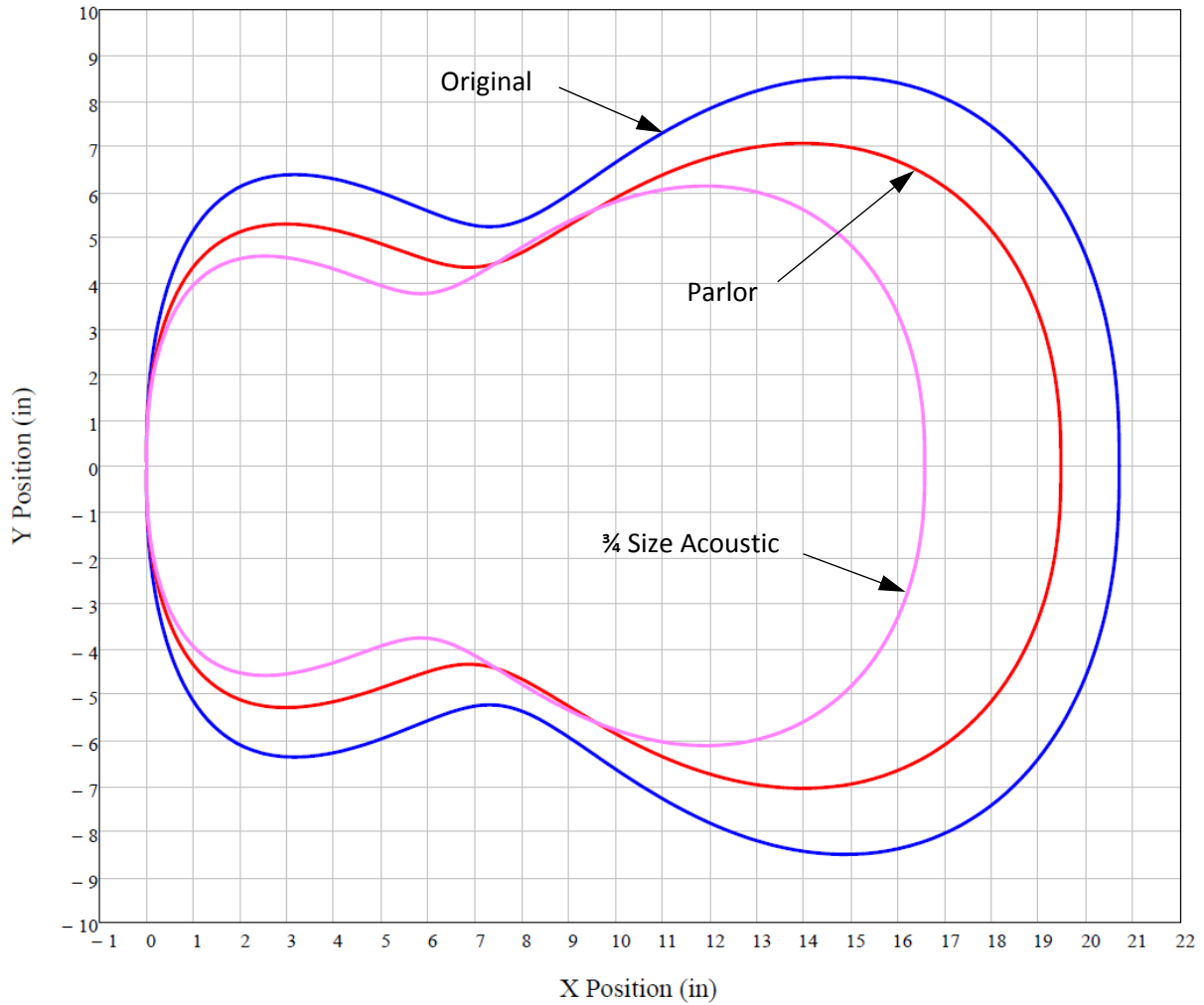

**Figure 10. Producing Different Size Bodies by Scaling the Basic Shape.**

Table 2 shows the scaling factors used to produce the shapes in Figure 11. A family of guitars based on these body shapes is being developed. The early focus will be on the parlor and ¾ shapes, though a ukulele may be added.

**Table 2. Scale Factors for Different Body Shapes (dimensions in inches).**

| Name | Body Length (in) | Lower Bout Width (in) | X Scale Factor | Y Scale Factor |
|---|---|---|---|---|
| Original | 20.7 | 17 | 1 | 1 |
| Parlor | 19.5 | 14.1 | 0.94 | 0.83 |
| ¾ Acoustic | 16.5 | 12.25 | 0.8 | 0.72 |

**Acknowledgements**

The author wishes to thank R.M. Mottola for sharing his expertise in the design and classification of acoustic guitars and for suggesting improvements to this article.

**References**

1) Johnston, Richard and Boak, Dick. 2009. *Martin Guitars: A Technical Reference*. Hal Leonard.

2) Cumpiano, William and Natelson, J. 1994. *Guitarmaking*. Chronical Books

3) Bruné, Richard. 2004. "Eight Concerns of Highly Successful Guitar Makers". *American Lutherie* 79:6-21.

4) Kreyszig, Erwin. 2015. *Advanced Engineering Mathematics*, 11$^{th}$ ed., Wiley.

5) French, Mark. 2006. "A Different Way of Defining Body Shapes". *American Lutherie* 88:52-57.

6) "MATLAB Curve Fitting Toolbox". Accessed January 5 2017. www.mathworks.com/products/curvefitting.html